\documentclass[journal,12pt,onecolumn,draftclsnofoot,]{IEEEtran}
\usepackage[utf8]{inputenc}
\usepackage{multirow}
\usepackage{soul}
\usepackage{algorithm}
\usepackage{algpseudocode}
\usepackage{graphicx}
\usepackage{dsfont}
\usepackage{xcolor}

\usepackage[most]{tcolorbox}

\usepackage[utf8]{inputenc}
\usepackage{mathtools}
\usepackage[mathscr]{euscript}
\usepackage{amsmath}
\usepackage{amssymb}
\usepackage{amsfonts}
\usepackage{amssymb}
\usepackage{amsmath}
\usepackage{verbatim}
\usepackage[T1]{fontenc}
\usepackage[utf8]{inputenc}
\usepackage{latexsym}
\usepackage{enumerate}
\usepackage{indentfirst}
\usepackage{calligra}
\usepackage{ulem}
\usepackage{graphicx}
\usepackage{array}
\usepackage{float}
\usepackage{bbm}

\usepackage{mleftright}
\usepackage{colortbl}
\makeatletter
\let\old@ps@headings\ps@headings
\let\old@ps@IEEEtitlepagestyle\ps@IEEEtitlepagestyle
\def\psccfooter#1{%
 \def\ps@headings{%
 \old@ps@headings%
 \def\@oddfoot{\strut\hfill#1\hfill\strut}%
 \def\@evenfoot{\strut\hfill#1\hfill\strut}%
 }%
 \def\ps@IEEEtitlepagestyle{%
 \old@ps@IEEEtitlepagestyle%
 \def\@oddfoot{\strut\hfill#1\hfill\strut}%
 \def\@evenfoot{\strut\hfill#1\hfill\strut}%
 }%
 \ps@headings%
}
\makeatother

\usepackage{soul}
\usepackage{multirow}
\usepackage{soul,color}
\usepackage{graphicx}
\usepackage{dsfont}
\usepackage{subcaption}

\usepackage[utf8]{inputenc}
\usepackage{mathtools}
\usepackage[mathscr]{euscript}
\usepackage{amsmath}
\usepackage{amssymb}
\usepackage{amsfonts}
\usepackage{verbatim}
\usepackage[T1]{fontenc}
\usepackage[utf8]{inputenc}

\usepackage{latexsym}
\usepackage{enumerate}
\usepackage{indentfirst}
\usepackage{calligra}
\usepackage{ulem}
\usepackage{multirow}

\usepackage{array}
\usepackage{float}
\usepackage{bbm}

\usepackage{eurosym}

\usepackage{hyperref}

\usepackage{caption}
\usepackage{tikz}
\usepackage{pgfplots}

\pgfplotsset{compat=1.8}
\usepgfplotslibrary{statistics}
\pgfmathdeclarefunction{fpumod}{2}{%
 \pgfmathfloatdivide{#1}{#2}%
 \pgfmathfloatint{\pgfmathresult}%
 \pgfmathfloatmultiply{\pgfmathresult}{#2}%
 \pgfmathfloatsubtract{#1}{\pgfmathresult}%
 \pgfmathfloatifapproxequalrel{\pgfmathresult}{#2}{\def\pgfmathresult{5}}{}%
 }

\usepackage{tabularx}
\usepackage{textcomp}
\usepackage{xcolor}
\usepackage{import}

\usepackage{csvsimple}
\usepackage{tabularx}
\usepackage{adjustbox}

\usetikzlibrary{trees,decorations,shadows}
\tikzset{level 1/.style={sibling angle=45,level distance=4mm}}
\usetikzlibrary{arrows.meta}
\usepackage{forest}
\usetikzlibrary{external}

\let\oldtikzexternalgetnextfilename\tikzexternalgetnextfilename \renewcommand{\tikzexternalgetnextfilename}[1]{\oldtikzexternalgetnextfilename{#1}\expandafter\tikzsetnextfilename\expandafter{#1}}

\usepackage{pgfplotstable}
\usepackage[outline]{contour}
\contourlength{1.2pt}

\pgfplotsset{compat=1.13} 

\usetikzlibrary{spy}
\usetikzlibrary{calc}
\usetikzlibrary{fadings}
\usetikzlibrary{patterns}
\usetikzlibrary{shadows}
\usetikzlibrary{mindmap}
\usetikzlibrary{backgrounds}
\usetikzlibrary{shapes.symbols}
\usetikzlibrary{shapes.multipart}
\usetikzlibrary{shapes.geometric}
\usetikzlibrary{automata,positioning}
\usetikzlibrary{decorations.fractals} 
\usetikzlibrary{decorations.markings}
\usetikzlibrary{decorations.pathreplacing}
\usetikzlibrary{decorations.pathmorphing}

\usepackage{nomencl}
\makenomenclature
\tikzset{edge from parent/.style={segment angle=10,draw}}

\tikzset{
 my rounded corners/.append style={rounded corners=2pt},
}

\def\BibTeX{{\rm B\kern-.05em{\sc i\kern-.025em b}\kern-.08em
 T\kern-.1667em\lower.7ex\hbox{E}\kern-.125emX}}

\renewcommand{\nomgroup}[1]{%
 \ifthenelse{\equal{#1}{O}}{\item[\textit{Operators}]}{%
 \ifthenelse{\equal{#1}{I}}{\item[\textit{Indices}]}{%
 \ifthenelse{\equal{#1}{A}}{\item[\textit{Acronyms}]}{%
 `\ifthenelse{\equal{#1}{V}}{\item[\textit{Variables and parameters}]}{}}}}}
\usepackage{scalerel}
\usetikzlibrary{svg.path}
\definecolor{orcidlogocol}{HTML}{A6CE39}
\tikzset{
 orcidlogo/.pic={
 \fill[orcidlogocol] svg{M256,128c0,70.7-57.3,128-128,128C57.3,256,0,198.7,0,128C0,57.3,57.3,0,128,0C198.7,0,256,57.3,256,128z};
 \fill[white] svg{M86.3,186.2H70.9V79.1h15.4v48.4V186.2z}
 svg{M108.9,79.1h41.6c39.6,0,57,28.3,57,53.6c0,27.5-21.5,53.6-56.8,53.6h-41.8V79.1z M124.3,172.4h24.5c34.9,0,42.9-26.5,42.9-39.7c0-21.5-13.7-39.7-43.7-39.7h-23.7V172.4z}
 svg{M88.7,56.8c0,5.5-4.5,10.1-10.1,10.1c-5.6,0-10.1-4.6-10.1-10.1c0-5.6,4.5-10.1,10.1-10.1C84.2,46.7,88.7,51.3,88.7,56.8z};
 }
}

\newcommand\orcidicon[1]{\href{https://orcid.org/#1}{\mbox{\scalerel*{ \begin{tikzpicture}[yscale=-1,transform shape]
 \pic{orcidlogo};
 \end{tikzpicture}
 }{|}}}}

\begin{document}
%
\title{DER Hosting capacity for distribution networks: definitions, attributes, use-cases and challenges }

 


\author{\IEEEauthorblockN{Md. Umar Hashmi}
 
 \IEEEauthorblockA{\textit{KU Leuven \& EnergyVille},
Genk, Belgium}

 \IEEEauthorblockA{mdumar.hashmi@kuleuven.be}
 }


\maketitle




\begin{abstract}
\small{
    The rapid adoption of distributed energy resources (DERs) has outpaced grid modernization, leading to capacity limitations that challenge their further integration. Hosting Capacity Assessment (HCA) is a critical tool for evaluating how much DER capacity a grid can handle without breaching operational limits. HCA serves multiple goals: enabling higher DER penetration, accelerating grid connection times, guiding infrastructure upgrades or flexible resource deployment, and ensuring equitable policies. 
HCA lacks a universal definition, due to varying modelling approaches, uncertainty considerations, and objectives. This paper addresses five key questions to standardize  HCA practices and applications. 
First, it classifies HCA objectives associated with different stakeholders such as system operators, consumers, market operators and consumers.
Second, it examines model attributes, including modelling sophistication, data requirements, and uncertainty handling, thus balancing complexity with computational efficiency. Third, it explores HCA applications, such as planning grid investments or operational decisions, and summarizes use cases associated with HCA. Fourth, it emphasizes the need for periodic updates to reflect dynamic grid conditions, evolving technologies, and new DER installations. Finally, this work identifies challenges, such as ensuring data quality, managing computational demands, and aligning short-term and long-term goals.
By addressing these aspects, this paper provides a structured approach to perform and apply HCA, offering insights for engineers, planners, and policymakers to manage DER integration effectively.}
\end{abstract}


\pagebreak

\tableofcontents

\pagebreak

\section{Introduction}
\vspace{-7pt}
The modernization of power grids has lagged behind the rapid adoption of distributed energy resources (DERs). As a result, many power networks face capacity limitations (manifesting as congestion and deteriorating power quality incidents) that hinder the efficient integration of DERs\footnote{\footnotesize{
DERs are assets, typically less than 10 MW, that can either generate, consume or store energy.
The DERs are typically located close to the sites where electricity is used \cite{ieaDER}.
The most common examples of DERs include photovoltaic (PV) systems, electric vehicles (EVs), batteries, heat pumps (HPs) etc \cite{gridX}.
In contrast to conventional generation and demand, DERs are modular and can be operated a decentralized manner, offering greater temporal flexibility.
}
}.
A systematic assessment known as Hosting Capacity Assessment (HCA) has become crucial to address this issue. HCA aims to evaluate how much DER capacity a power grid can accommodate without violating operational constraints. It serves several important objectives, including (\textit{i}) facilitating the integration of higher levels of DERs, (\textit{ii}) reducing the time required for DERs to connect to the grid, (\textit{iii}) assisting in the planning of network upgrades or the prioritization of flexible energy resources, (\textit{iv}) ensuring equitable policy frameworks for all stakeholders, and (\textit{v}) enhancing grid operational flexibility while minimizing curtailment. 


Distribution networks (DNs) have limited capacity, and rising DER integration is straining the power networks. Grid capacity constraints already hinder DER deployment
and new consumer connections. Inefficient connection procedures exacerbate the problem, leading to long delays and risks of stalling the ongoing energy transition.
Hosting Capacity Analysis (HCA) has emerged as a critical tool to ensure reliable and efficient grid operation amidst this transformation. HCA assesses the ability of power grids to accommodate DERs without exceeding operational constraints such as voltage limits, thermal capacity, power quality limits and protection coordination. This paper delves into the complexities of DER HC by addressing five pivotal questions, each aimed at advancing the understanding and application of HCA.

Despite its significance, HCA lacks a universally accepted definition. This is primarily because HCA varies based on factors like the level of detail used in modelling, how uncertainties are factored into the assessment, and the specific objectives driving the need for an HCA study. In light of these differences, this paper seeks to answer several key questions to help standardize the understanding and application of HCA in power systems. First, we seek to classify HCA based on its various objectives. Depending on the context, HCA may be used to assess the grid's ability to host DERs under different scenarios, such as normal operating conditions or peak demand. In other cases, HCA may focus on identifying where grid upgrades are most needed or where operational flexibility can be maximized. 
The variation in objectives shapes the scope and approach of HCA studies, impacting their relevance for DER owners, planners, operators, and policymakers.
Second, the paper explores the attributes of models used for HCA, such as the level of modelling detail, data requirements, and methods to address uncertainties. These attributes significantly influence the accuracy, computational speed and applicability of HCA results.
Third, the paper explores the different use cases of HCA. These may range from planning purposes, such as determining where to invest in grid upgrades, to real-time operational decisions, like managing DER curtailment during extreme operating cases. Understanding these diverse use cases is essential for tailoring HCA studies to meet the specific needs of power system operators, planners, policymakers and other stakeholders. Fourth, we address the question of how frequently HCA should be conducted. Power networks are dynamic, with changing load patterns, new DER installations, and evolving technologies. As such, regular updates to HCA are necessary to ensure that the grid can continue to accommodate DERs safely and efficiently. Finally, we outline the challenges associated with performing HCA. These challenges include the need for high-quality data, the computational complexity of HCA models, and the difficulty of balancing short-term operational needs with long-term planning goals. 




In summary, by addressing these five key questions, we provide a comprehensive understanding of the HCA problem. Through this structured approach, we offer valuable insights that will help power system engineers, researchers, planners, DER/flexibility owners, system operators, market operators, and policymakers effectively manage the integration of DERs into the grid.


\pagebreak

\section{Hosting capacity: definitions and classifications}
Most HC definitions deal with distributed generation (DG), for instance, {EPRI definition} for HC states: 
"\textit{Hosting capacity is defined as the amount of PV
that can be accommodated without impacting
power quality or reliability under existing control
and infrastructure configurations}" \cite{rylander2015distribution}.

\begin{tcolorbox}[colback=gray!8!white,colframe=blue!75!black]
\small{
\textbf{Rule of thumb based HC estimates}: frequently used estimates for DG HC based on customer demand are 15\% of the peak demand and/or 100\% of the minimum demand
\cite{lindl2013integrated}. However, such estimates are more difficult to design for DERs that increase the load on the feeder/network such as EVs, HPs etc.
}
\end{tcolorbox}

\begin{figure}[!htbp]
    \centering
    \includegraphics[width=6.2in]{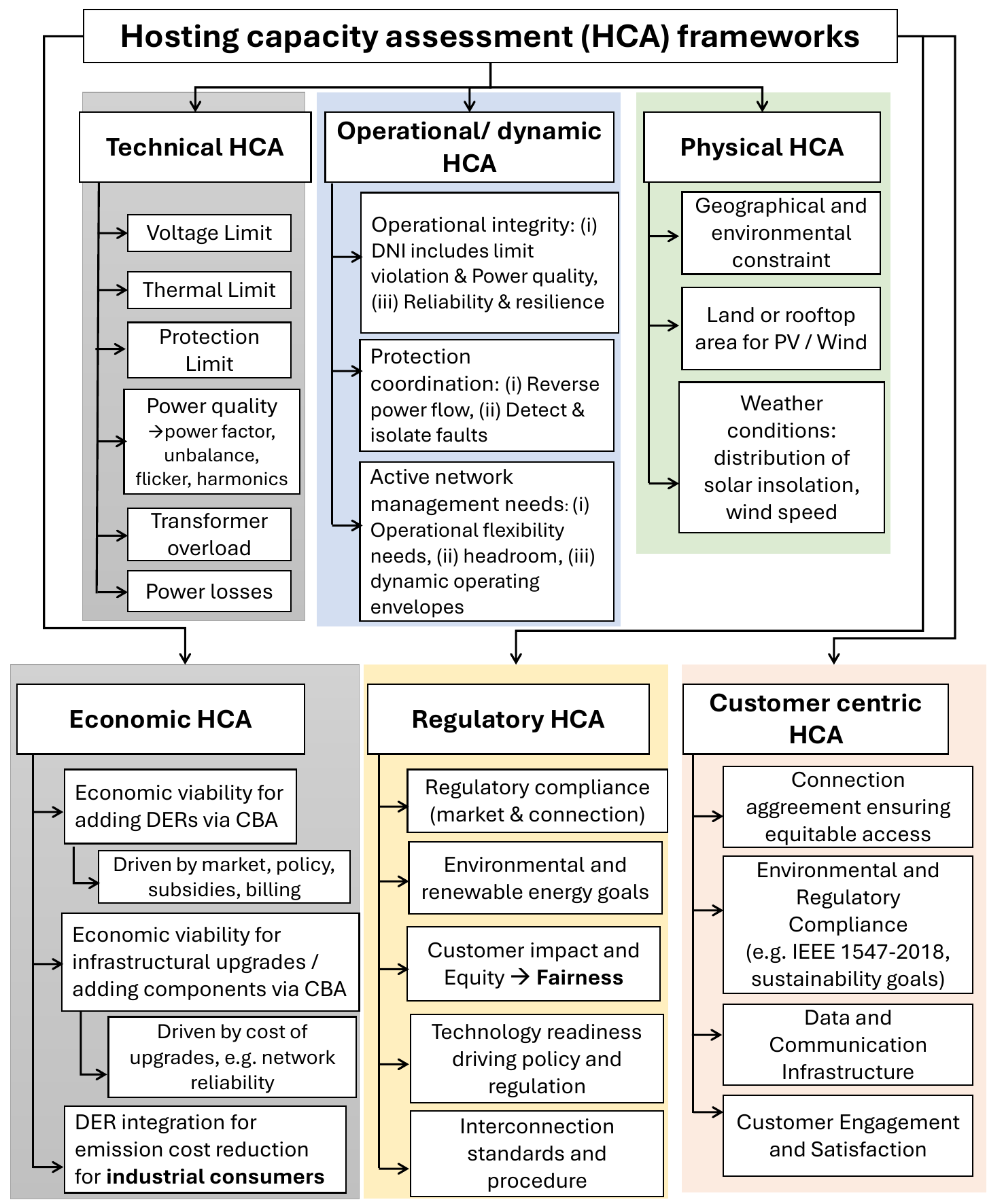}
    \vspace{-8pt}
    \caption{\small{Definitions for hosting capacity based on the objectives.}}
    \label{fig:hostingTypes}
\end{figure}

The general definition utilized in this work is "\textit{the ability for the network to accommodate a specific or combined installed capacity of one or multiple DER technology considering technical, operational, economical, physical, regulatory, equitable and dynamic considerations on a specific circuit without negatively affecting reliability or requiring infrastructure upgrades}". However, HCA can be classified based on the different objectives that are aimed at depending on timescale/horizon, stakeholder's local objectives and the type of analysis such as economic cost-benefit analysis is crucial for investment decisions, but technical network limitations are necessary for reliable operation of the power network. Next, we provide six different HCA frameworks as shown in Fig. \ref{fig:hostingTypes}.


\vspace{-10pt}

\pagebreak

\subsection{Evolving realms of hosting capacity}
Hosting capacity is an evolving domain of research in the field of power systems. The definitions proposed earlier do not reflect the nuanced utility of this tool for different entities in power networks. 
Instead of a single definition for HCA, we propose 6 different definitions based on the objectives and stakeholders.

Table \ref{tab:hcatypeStake} details the different stakeholders who may be performing the different adaptations of HCA.

$\bullet$ \textit{Technical HCA} (T-HCA) is the maximum DERs a power DN can support reliably and efficiently without major upgrades, considering technical constraints and power quality standards.
\vspace{10pt}


$\bullet$ \textit{Dynamic HCA} (D-HCA) is the maximum amount of DERs that a DN can accommodate at any given time while maintaining stable and reliable operation in real-time. Unlike static HC, which is a fixed value based on worst-case scenarios or average conditions, dynamic HC considers variability and fluctuations in both generation and load. Thus, dynamic HC is a time-varying attribute.
\vspace{10pt}

$\bullet$ \textit{Physical HCA} (P-HCA) refers to the maximum potential for integrating DERs into the electrical grid, limited by geographical, weather, and land-use constraints. Unlike technical HC, which is governed by technical factors like power quality limits, physical HC is determined by the physical environment, such as rooftop area availability for residential PV installations in urban spaces.
\vspace{10pt}


$\bullet$ \textit{Economic HCA} (E-HCA) refers to the economic viability assessment for DER integration from the system operator, DER owner, aggregator perspectives via cost benefit analysis (CBA). E-HCA is driven by market rules, subsidies, billing mechanisms, cost of network upgrades, and the cost of DER installations. Industrial consumers while assessing E-HCA also identifies the emission cost reduction for considering all revenue streams.
\vspace{10pt}

$\bullet$ \textit{Regulatory and policy-based HCA} (R-HCA) refers to the limits or guidelines set by regulatory authorities and policy frameworks that govern how much DERs can be connected to the electricity DN. These limits are not solely based on technical considerations like thermal limits, voltage regulation, or stability but are also influenced by regulatory decisions, policy objectives (fairness), and social or economic factors. 
\vspace{10pt}

\begin{table}[!htbp]
\centering
\small
\caption{\small{HCA type for different stakeholders}}
\begin{tabular}{p{25mm}|p{12mm}|p{47mm}} 
\hline
HCA type                   & Short form & Stakeholder                                                                  \\ 
\hline \hline
Technical HCA              & T-HCA      & Distribution System operator (DSO)                                                             \\ 
\hline
Dynamic HCA & D-HCA      & DSO, aggregator, community manager (CM)                              \\ 
\hline
Physical HCA               & P-HCA      & DSO, policy maker                                                \\ 
\hline
Economic HCA  & E-HCA      & DSO, aggregator, CM, prosumer\\ 
\hline
Regulatory HCA  & R-HCA      & Regulator, policy maker, market operator                                     \\ 
\hline
Consumer-centric HCA       & C-HCA      & Prosumer, flexibility owner, CM                              \\
\hline
\end{tabular}
\label{tab:hcatypeStake}
\end{table}


\begin{tcolorbox}[colback=gray!8!white,colframe=blue!75!black]
\small{
\textbf{Feasibility studies}: DSOs perform feasibility studies to determine if the network can support a proposed connection without breaching operational limits or requiring major upgrades. They analyze power flows, voltage profiles, congestion, power quality, and protection system adequacy to ensure fault detection and coordination remain intact. The study also evaluates short-circuit current handling, upgrade costs versus DER benefits, and compliance with grid codes and interconnection standards. 
}
\end{tcolorbox}

$\bullet$ \textit{Consumer-centric HCA} refers to the ability of a power DN to integrate DERs in a manner that prioritizes the needs and preferences of consumers. This approach goes beyond the traditional technical constraints and focuses on how the DN can support consumer-driven energy choices and equitable opportunities to install DERs while maintaining reliability, and power quality.

\pagebreak
\subsection{DER adoption vs HC of DN}
\begin{figure}[!htbp]
    \centering
    \includegraphics[width=0.75\linewidth]{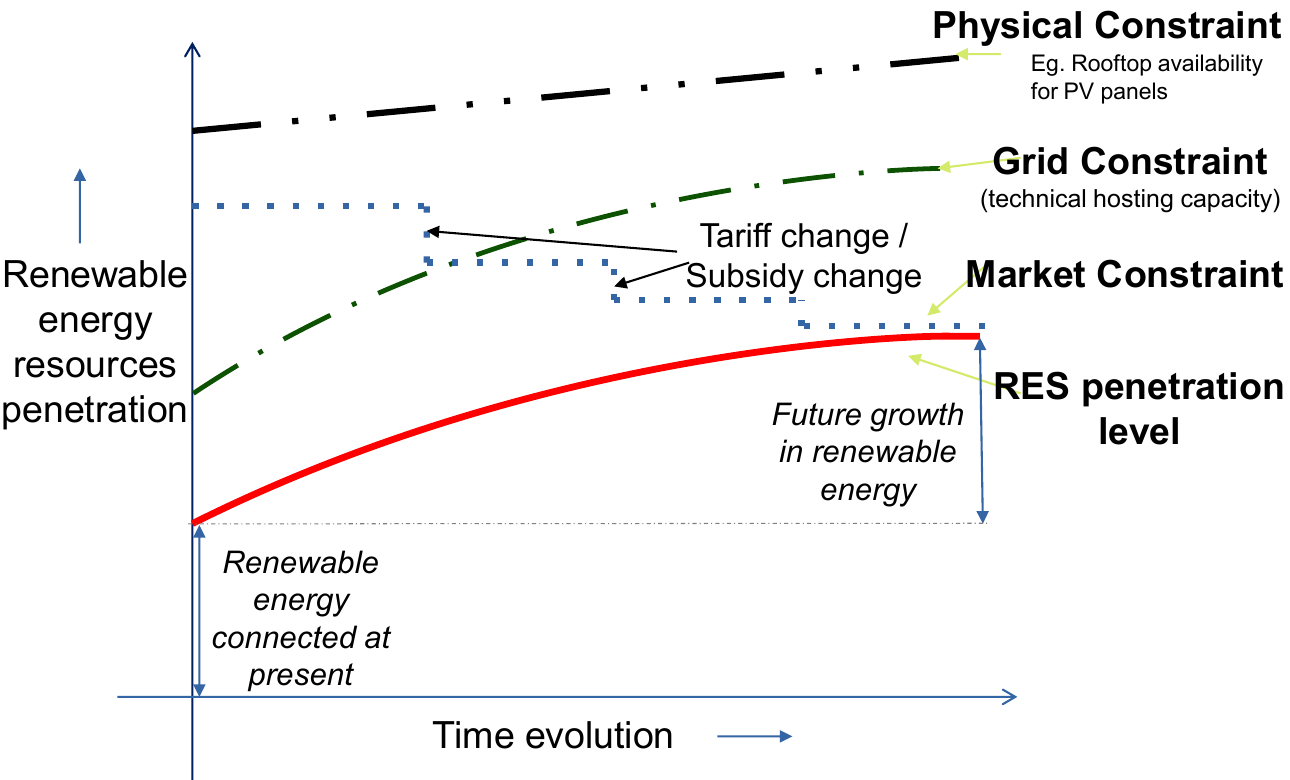}
    \vspace{-9pt}
    \caption{\small{Stylized example showcasing physical, grid and market constraints affecting RES adoption}}
    \label{fig:resadoption}
\end{figure}

Although the technical HCA provides us with details of how much DER can be connected, the evolution of the penetration of DER is not just a function of technical HCA, but also the physical constraint detailed by physical HCA, financial viability aspects covered within economic HCA. The evolution of RES penetration based on stylized consumers that shows the evolution in time with different constraints is shown in Fig. \ref{fig:resadoption}.

\begin{tcolorbox}[colback=gray!8!white,colframe=blue!75!black]
\small{
\textbf{Injection limits \& reactive power}:
In Germany, inverter sizing for PV systems is shaped by the "70\% rule," limiting maximum feed-in power to 70\% of the installed capacity. This ensures grid stability, minimizes network strain, and encourages self-consumption of generated electricity while reducing the need for grid upgrades.
Inverters must also meet reactive power requirements, supporting specific power factors (0.9–0.95, lead/ lag) for voltage support \cite{ratedpower}. 
}
\end{tcolorbox}

\pagebreak

\section{Attributes of the models for HCA}
\begin{figure}[!htbp]
    \centering
    \includegraphics[width=0.899\linewidth]{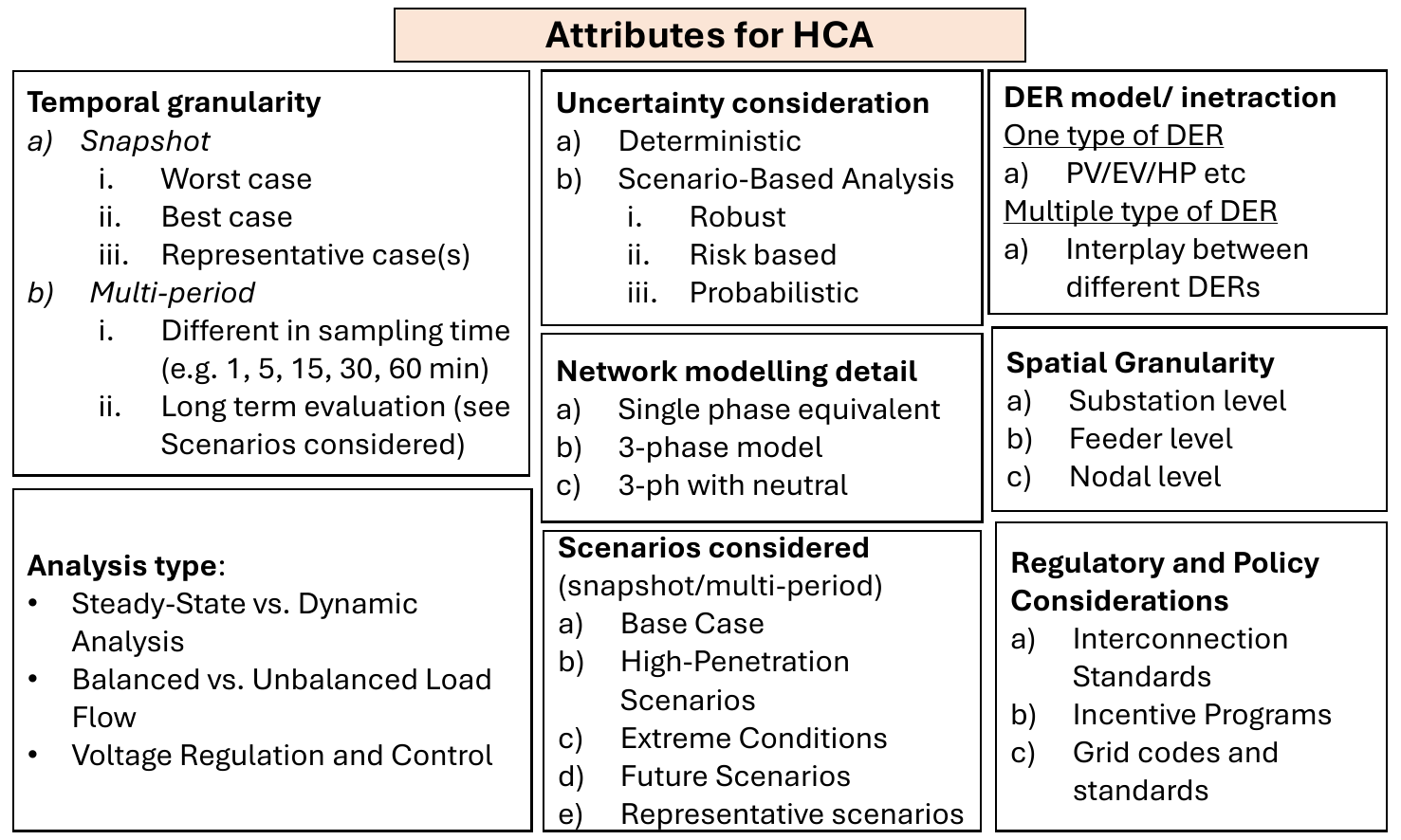}
    \caption{\small{Attributes of HCA models}}
    \label{fig:hostingAttributes}
\end{figure}

\vspace{-5pt}
The attributes of HCA encompass a wide range of technical, regulatory, and modelling considerations to ensure ballpark evaluations of the DN's ability to integrate DERs, see Fig. \ref{fig:hostingAttributes}.

.

The \textit{analysis type} determines the approach and scope of the study. It can involve steady-state versus dynamic analysis, focusing on either the static conditions or the time-varying behavior of the network. The load flow assessment may be balanced or unbalanced, accounting for asymmetries that are often present in real-world scenarios. Voltage regulation and control are critical components, ensuring that voltage levels remain within permissible limits under various operating conditions.

\textit{Regulatory and policy considerations} are integral to HCA. 
Grid codes and standards provide the overarching framework to ensure operational reliability, safety, and compliance with national and regional policies.
Interconnection standards dictate the technical requirements for connecting DERs to the grid (e.g. IEEE-1547-2018\footnote{IEEE 1547-2018 is a standard governing interconnection and interoperability requirements for DERs within electric power systems.}), while incentive programs encourage investments in DERs. 

The \textit{detail of the network modelling} significantly impacts the analysis. For instance, HCA may involve a single-phase equivalent model for simplified representations, a full three-phase model for more detailed and accurate studies, or a three-phase model with neutral to account for neutral current flows and imbalances in the DN. These modelling details lead to a more accurate HC value, at the cost of greater computational needs.


\begin{tcolorbox}[colback=gray!8!white,colframe=blue!75!black]
\small{\textbf{Connection limits}: 
Single-phase systems are more commonly restricted due to the risk of voltage unbalances, especially in areas with high DER penetration.
Three-phase systems offer higher capacity limits and better distribution of load across phases, reducing grid stress.
Typical limits include \cite{horowitz2019overview,lucas2018single}:
(i) 3.68 kW in many countries like Austria, Denmark, and Estonia.
(ii) 4.6 kW in Germany and Poland.
(iii) 5–6 kW in countries like France and Greece.
(iv) For three-phase systems, capacities can go higher, such as 10 kW in Romania and up to 30 kW in Luxembourg
(v) Australia commonly caps single-phase PV systems at 5 kW for residential installations. Three-phase connections can support up to 15 kW, but the limits may vary depending on the local grid and consideration of voltage rise.
(vi) In the U.S., DER integration capacity is guided by standards like IEEE 1547. The limit for individual DER systems is generally tied to the HC of the local DN, with specific restrictions varying by utility. 
For EVs, single-phase chargers typically operate at 1.6 to 7.4 kW for residential and small commercial installations. For example, Type 2 chargers in Europe provide up to 7.4 kW on a single-phase connection. In North America, single-phase systems using SAE J1772 Type 1 can deliver up to 7.7 kW \cite{duevolt}.
}
\end{tcolorbox}

\textit{DER modelling and interaction} can focus on a single type of DER, such as PV, EV, or HP. Alternatively, it may explore the interplay between multiple types of DERs, highlighting their combined impacts on the DN and the associated integration challenges.

\textit{Spatial granularity} defines the level of detail at which the network is analyzed. It may be at the substation level for broad insights, the feeder level for medium-granularity studies, or the nodal level for detailed evaluations of specific network locations.

\textit{Temporal granularity} addresses the time dimension of the analysis. A snapshot approach evaluates specific instances, such as worst-case, best-case, or representative scenarios. Multi-period analysis provides a more dynamic perspective, using various sampling intervals (e.g., 1, 15, or 60 minutes) to capture time-dependent variations. Long-term evaluations may consider scenarios over months or years to predict future grid behavior.

\textit{Uncertainty considerations} play a crucial role in managing the unpredictability inherent in DER integration. The analysis can be deterministic for fixed conditions or involve scenario-based approaches to explore different potential futures. Robust, risk-based, and probabilistic methods provide a deeper understanding of uncertainties, ensuring more resilient planning and operation.

\begin{figure}[!htbp]
    \centering
    \includegraphics[width=0.85\linewidth]{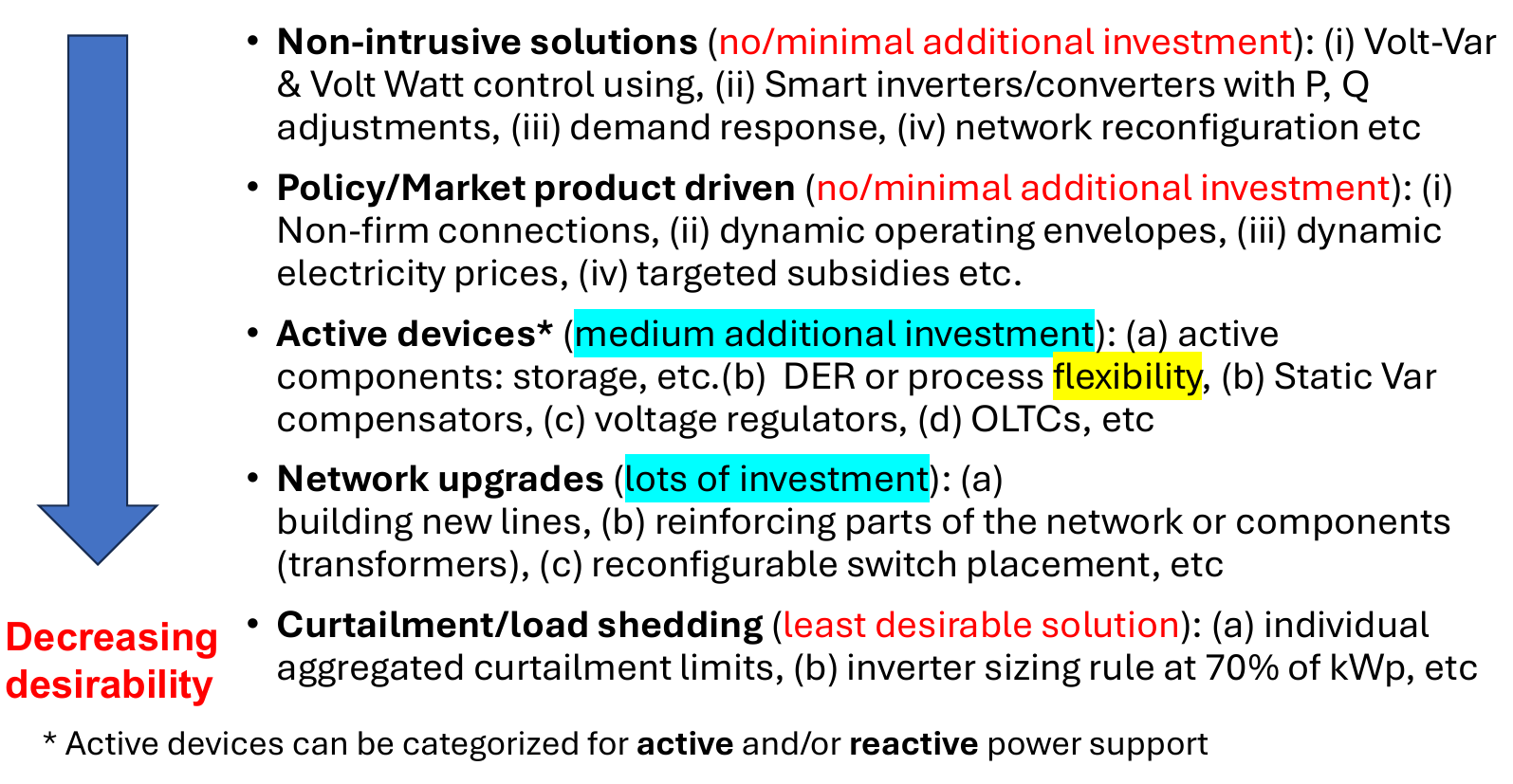}
    \caption{\small{DER hosting capacity improvement techniques}}
    \label{fig:HCAimprovement}
\end{figure}

Finally, the \textit{scenarios considered} offer diverse perspectives to accommodate various possible grid conditions. These may include a base case for current operations, high-penetration scenarios for increased DER integration, extreme conditions to test system limits, future scenarios for long-term planning, and representative scenarios to capture typical network states.

From Fig. \ref{fig:hostingAttributes}, adding modelling sophistication would often lead to increased computational needs for solving the HCA problem. The concerned stakeholders should consider the modelling detail needed for their specific objectives.




\textit{HC improvement techniques}
There is a huge pool of literature on frameworks for improving network HC. Fig. \ref{fig:HCAimprovement} summarizes the HC improvement techniques.

\pagebreak
\section{Goals \& Use cases of HCA}
\begin{figure}[!htbp]
    \centering
    \includegraphics[width=0.899\linewidth]{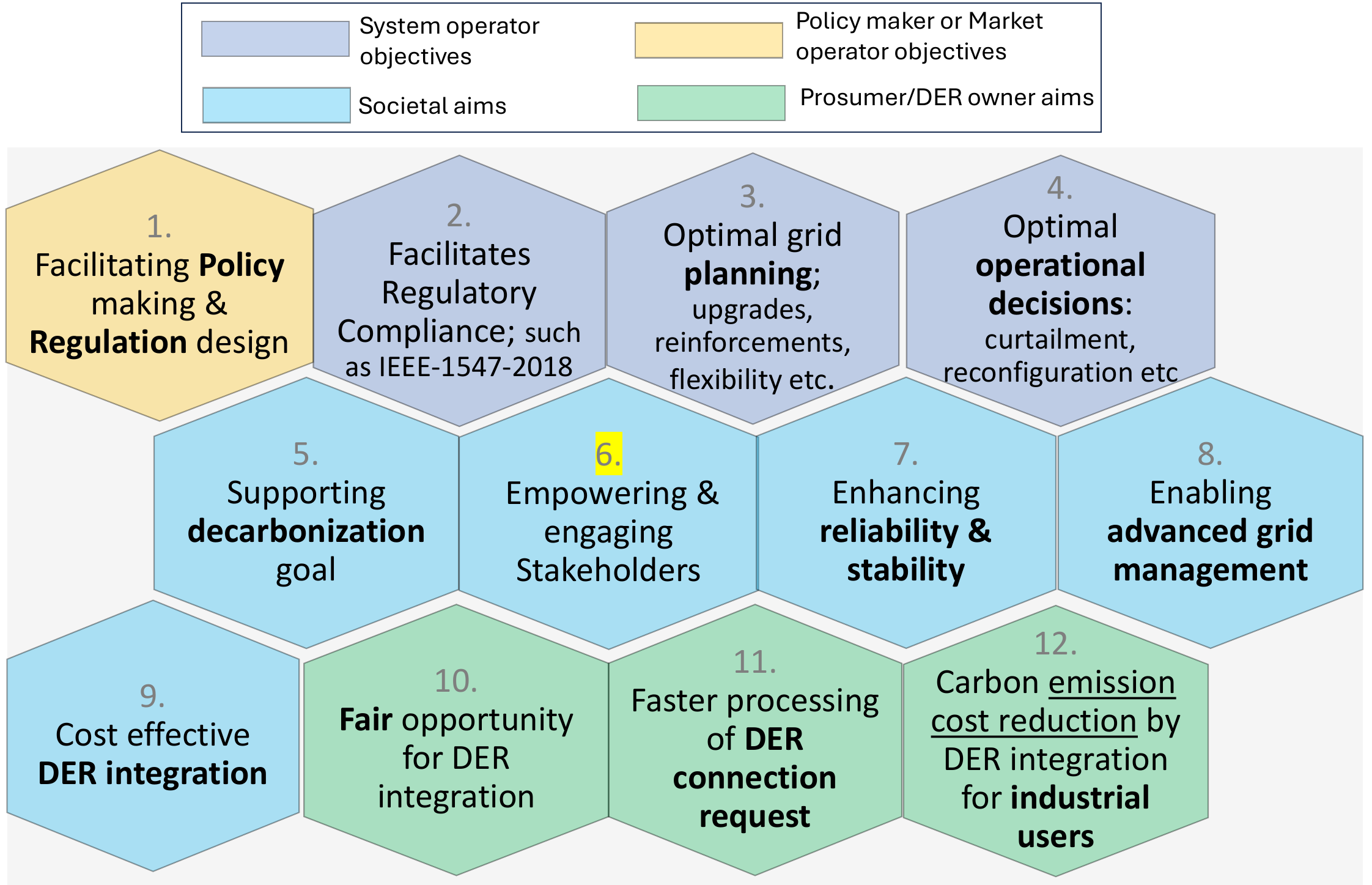}
    \vspace{-8pt}
    \caption{\small{The different use cases for performing HCA of power networks}}
    \label{fig:hostingUsefulness}
\end{figure}

\vspace{-7pt}
As noted above, HCA is performed for multiple goals that affect multiple stakeholders. Without distinguishing between
T-HCA, D-HCA, P-HCA, E-HCA, R-HCA and C-HCA, the broad goals of performing HCA are detailed in Fig. \ref{fig:hostingUsefulness}. These objectives are categorized into 4 types: (i) system operator goals, (ii) prosumer/DER owner goals, (iii) policy maker or market operator objectives and (iv) societal goals.
The first three goals should ideally lead to societal goals, however, we explicitly detail them here.

The stakeholder impacts/utilities are summarized below:\\
$\bullet$ \textit{For DSOs}: DER HC provides DSOs with crucial information to manage the grid effectively. It aids in maintaining grid reliability, optimizing grid operations, and avoiding costly and reactive network upgrades. This also supports DSOs in meeting regulatory requirements for grid stability and reliability.\\
$\bullet$ \textit{For Aggregators}: 
Aggregators managing DER portfolios can use HC data to strategically deploy and operate resources. Knowing HC across the grid helps optimize DER placement, maximize market participation returns, and offer more reliable services to the grid, such as demand response and ancillary services.\\
$\bullet$ \textit{For Consumers}: Consumers, particularly those investing in DERs (like rooftop solar, EV, HP), benefit from understanding the HC of their local network. This information helps them make informed decisions about the size and type of DER they can install without risking grid-related issues like frequent curtailment. Consumers can also engage more effectively in local energy markets, potentially earning income through services such as demand response, energy export, or P2P trading.
\\
$\bullet$ \textit{For market operators}: HCA supports policymakers and market operators by providing data-driven insights critical for crafting equitable policies and regulations for DNs. It enables the design of frameworks that balance the needs of all actors, including system operators, consumers, and aggregators, ensuring fair access and participation in the energy market. By identifying technical and operational constraints, HCA helps policymakers address grid challenges through targeted interventions and incentives, fostering a more inclusive and competitive market. 
\\
$\bullet$ \textit{For societal benefits}: 
HCA fosters a collaborative, efficient, and sustainable energy system, empowering stakeholders like DSOs, consumers, and market operators with transparent grid insights for informed decisions. It enhances grid reliability by identifying issues proactively, prevents outages, and supports cost-effective DER integration by optimizing infrastructure investments. HCA also addresses technical and economic barriers, enabling smarter, resilient grids to meet dynamic energy demands. By accelerating clean energy integration and reducing emissions, HCA drives decarbonization and supports a sustainable energy future, creating a more inclusive and reliable energy ecosystem.


\vspace{-5pt}
\subsection{Use case 1: Network inadequacy}
\vspace{-5pt}
Network inadequacy causes \textit{connection delays} for new/upgrading existing supply and demand requests: although the connection delays are governed by many causes such as permitting, material bottlenecks, digitalization, implication assessment etc. In this work, we will focus only on the connection delays caused due to feasibility studies necessary for incoming DERs. Such rules are often exempted for small DERs and/or in over-designed DNs with low penetration of DERs, however, as the share of DERs increases in the DNs administered by such DSOs, the validation of an incoming DER via feasibility study becomes mandatory, thus causing connection delays. 
HCA can be considered a feasibility study performed preemptively, thus, reducing connection delays.

\begin{tcolorbox}[colback=gray!8!white,colframe=blue!75!black]
\small{
\textbf{Connection delays}:
An incoming DER installation (greater than a threshold size DER set by DSO) needs to place a connection request. As a follow-up, the DSO needs to perform feasibility studies, leading to delays as often the processing of such connection requests may also be preceded by local network upgrades ensuring the operational integrity of the power networks. 
}
\end{tcolorbox}

 $\bullet$ For instance, network inadequacy leads to global connection delays, causing nearly 1,000 GW of solar projects—almost four times the amount of new solar capacity added worldwide in 2022—and 500 GW of wind projects to be stuck in the interconnection queue across the United States and Europe
 \cite{bnef_web}.
        
\textit{Solution}: Updated DER HC maps on a regular basis could substantially decrease the connection delays otherwise observed.

\vspace{-20pt}
\begin{figure}[!htbp]
    \centering
    \includegraphics[width=0.999\linewidth]{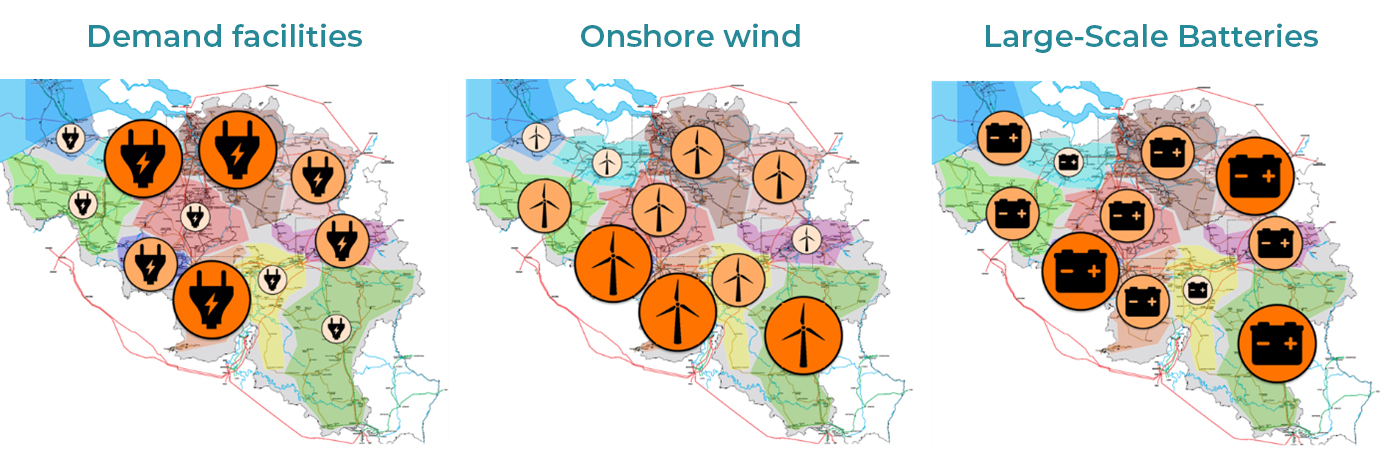}
    \vspace{-10pt}
    \caption{\small{DER hosting capacity visualization for Belgium: additional load, onshore wind and energy storage \cite{elia_hc}.}}
    \label{fig:belgiumHCA}
\end{figure}
\vspace{-19pt}
\vspace{-5pt}
\subsection{Use case 2: Hosting capacity maps}
\vspace{-5pt}
HCA is a valuable tool for evaluating the locational value of DERs as their presence on the grid increases. By using HC maps, which visually represent the outcomes of an HCA, information can be shared transparently among regulators, developers, electric customers, and utilities. This transparency fosters more efficient and cost-effective deployment of DERs across the grid.
{HC maps} are visual representations of an electrical grid’s ability to integrate DERs without compromising grid reliability or violating operational limits. These maps provide spatial insights into how much additional DER capacity can be safely accommodated at various locations within the grid, considering factors such as voltage regulation, thermal capacity, and protection coordination.
These maps are typically color-coded or layered with data overlays, showing varying levels of HC across substations, feeders, or individual nodes. By offering a geographic perspective, they help stakeholders make informed decisions for DER siting and interconnection. Fig. \ref{fig:belgiumHCA} shows HC maps for additional demand, wind generation and batteries performed by Belgian TSO, Elia.

Examples of Hosting Capacity Map Implementation: (A) \textit{California}: Under the California Public Utilities Commission (CPUC) Rule 21, utilities like PG\&E, SCE, and SDG\&E are required to publish and maintain interactive hosting capacity maps, updated regularly \cite{CPUC2023}. (B)
\textit{New York}: The New York State Public Service Commission mandates hosting capacity analysis as part of the Distributed System Implementation Plans under the Reforming the Energy Vision (REV) initiative \cite{edison2023consolidated}. (C)
\textit{Europe}: In Germany and the UK, hosting capacity maps are being explored as part of the transition to smarter, decentralized grids.
(D) \textit{Virginia}: Dominion Energy utilizes the HC tool for identifying locations to install utility-scale (MW) and residential scale (kW) DERs. This tool is part of a 10-year grid transformation plan to facilitate energy transition in Virginia. These HC maps are updated at least quarterly. For utility HC maps, each connected branch is coloured according to available capacity. For residential HC maps, the transformers are the limiting factor coloured in the HC maps \cite{dominion}.

\begin{figure}[!htbp]
    \centering
    \includegraphics[width=0.99\linewidth]{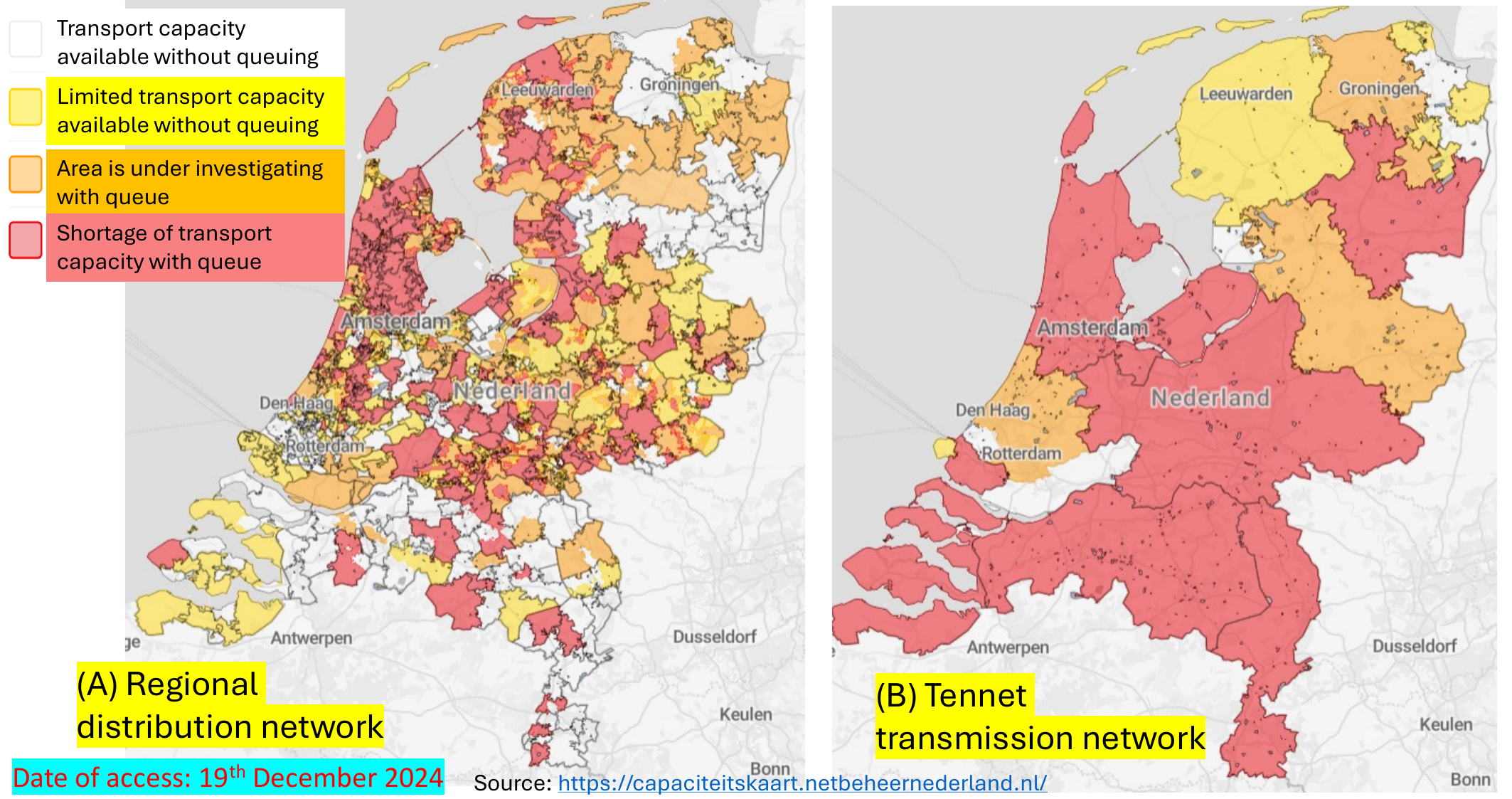}
    \caption{\small{DER hosting capacity visualization for the Netherlands: combining distribution and transmission \cite{netherlands}.}}
    \label{fig:dutch}
\end{figure}

In the Netherlands, HC maps more commonly referred to as capacity maps are already utilizing these maps for planning transmission grid expansions \cite{netherlands}, see Fig. \ref{fig:dutch}.

\pagebreak
\section{Frequency of performing HCA}
\vspace{-7pt}
The frequency of HCA for distribution networks can vary based on several factors. In many cases, there may be no specific policy for updating HCAs unless otherwise stipulated, leaving the process largely unregulated unless a formal mandate/policy exists.

\begin{figure}[!htbp]
    \centering
    \includegraphics[width=0.77\linewidth]{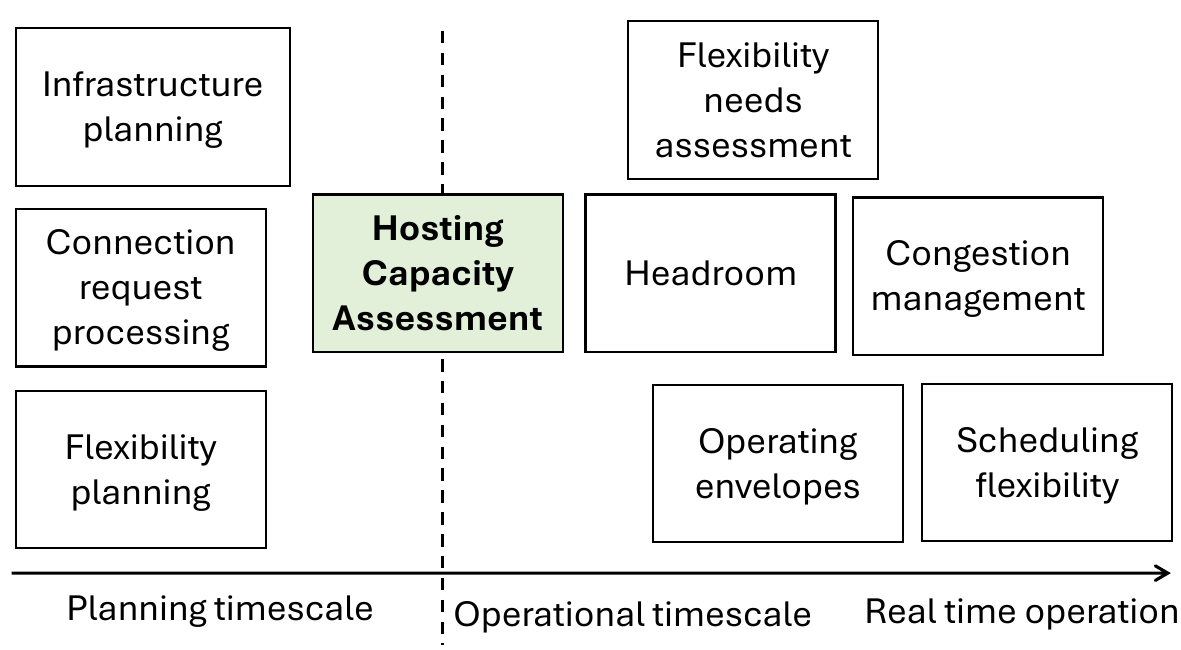}
    \vspace{-7pt}
    \caption{\small{Decision support tools for DN operation and planning.}}
    \label{fig:hostingPlanningOper}
\end{figure}

As evident from Fig. \ref{fig:hostingPlanningOper}
DN HCA is both a planning and operational problem. In planning, it evaluates the grid's ability to integrate DERs over time, guiding infrastructure upgrades, policy-making, and resource allocation. For operational purposes, it assesses real-time grid conditions, enabling network operators to make informed decisions about managing DERs, preventing congestion, power quality limit violation incidents, and maintaining system stability. By addressing both long-term and short-term needs, HCA ensures sustainable grid development and reliable day-to-day operation. 
One common approach is to update HCAs at fixed intervals, such as monthly, quarterly, or yearly. This ensures a regular review of the network's capacity to integrate DERs. However, some utilities prefer a more dynamic approach, updating HCAs when specific changes occur. These changes might involve network topology shifts/reconfiguration, upgrades to power lines, the connection of new DERs, or the inclusion of active network elements like on-load tap changers or voltage regulators. These updates are triggered when the change exceeds a predefined threshold, ensuring that HCAs reflect significant grid alterations.

In some regions, regulatory bodies dictate the frequency of HCA updates. For example, in California, the CPUC mandates that utilities perform and publish annual HCAs, with more frequent updates required in response to grid evolution or increasing DER penetration. Similarly, New York’s Public Service Commission requires regular updates as part of its Reforming the Energy Vision (REV) initiative. Countries like Germany and Denmark also have advanced frameworks for assessing HC, often aligning updates with the annual planning and investment cycles of DSOs. In contrast, other nations may conduct assessments less frequently due to slower adoption of DERs or different regulatory priorities.

Some advanced utilities are now moving toward real-time or near real-time HCA (D-HCA) updates, enabled by sophisticated grid management systems and real-time data analytics. This allows for dynamic updates as grid conditions change, providing more accurate and timely assessments of HC.

\pagebreak
\section{Challenges associated with HCA}
\vspace{-7pt}
Performing HCA across millions of DN feeders is daunting, particularly when considering the need for regular updates to accommodate the dynamic and stochastic nature of DERs. The sheer scale and frequency of these HCAs make it critical to find ways to \textit{significantly enhance computational performance} without compromising the utility of the HCA results. Given that the precise accuracy of HCA is less critical due to the discrete and ad hoc nature of DER integration and policy decisions, several strategic approaches can be employed to improve scalability and efficiency.

\textbf{Computational improvements} 
offer a direct path to enhancing HCA scalability. Traditional AC power flow calculations are computationally heavy, especially for many Monte Carlo scenarios. Data-driven or ML-based power flow models provide faster results, capturing power flow dynamics with reasonable accuracy. Physics-informed models, trained on historical data and simulations, enable rapid predictions, allowing quicker HCA updates.

\textbf{Handling uncertainty} is another critical aspect. 
Solely relying on deterministic or worst-case assessments may misjudge the network's HC, leading to suboptimal usage of DN assets. Monte Carlo simulations can address uncertainty but are computationally intensive \cite{fani2024impact}. Scenario reduction techniques mitigate this by focusing on critical scenarios, eliminating less impactful ones, and maintaining robust, efficient analyses.

\textbf{HCA scalability}: 
Simplifying the DN model can save computation time. Model order reduction reduces nodes, speeding up power flow analysis while preserving key characteristics. Efficiency improves further by clustering feeders into representative groups for HCA, allowing DSOs to extrapolate results across the network and cut down calculations \cite{koirala2022representative}.

\textbf{HCA standardization}:
In addition to computational improvements, standardization of the HCA approach across different DERs and networks is critical for ensuring consistency and scalability. A standardized methodology would streamline the process, allowing for easier comparison 
of results across different regions/ DSOs. 

\textbf{Data-intensive needs}:
HCA faces challenges in DNs due to the data-intensive framework and low system observability. Key issues include incorrect topology from outdated or incomplete network models and inaccurate load profiles, both critical for assessing DER integration. These data gaps hinder reliable HCA results, affecting voltage regulation and thermal constraints. Improved monitoring, data collection, and model validation are essential. Additionally, sparse DN measurements require robust methods like imputation, historical trend analysis, or proxy measurements to ensure reliable HCA despite incomplete data.


{\textbf{HC disclaimer}}: 
Many DSOs create HC maps, but disclaimers often limit their reliability, stating they are for informational purposes only and that DSOs are not liable for risks users incur. These maps rely on best-effort estimates, outdated data, and assumptions subject to change. Unauthorized reproduction or modification is prohibited \cite{dominion, fortisAlberta}. Such disclaimers deter broader adoption, requiring substantial efforts to build trust in HC map usage.


\textbf{Underlying assumptions}:
Almost all of the HCA methods consider prosumers to be passive entities in DN operation. This may not be a fair assumption in the near future.

In conclusion, addressing the scalability of the HCA problem in large, diverse DNs requires faster power flow calculations, scenario and model order reductions, and feeder clustering. Standardizing HCA methods, particularly for handling data scarcity and uncertainty, will boost efficiency and consistency. These strategies enable DSOs to assess and manage HC effectively, supporting DER integration while ensuring grid reliability.

\pagebreak
\section{Conclusions}
The integration of Distributed Energy Resources (DERs) into modern power grids presents both opportunities and challenges, particularly in the face of aging infrastructure and increasing energy demands. This paper emphasizes the critical role of Hosting Capacity Assessment (HCA) in overcoming these challenges by evaluating how much DER capacity the grid can support without breaching operational limits. By systematically addressing five key aspects—objectives, modelling attributes, use cases, frequency of assessment, and associated challenges—this study advances the understanding and application of HCA for different stakeholders in power distribution networks.

\section*{Acknowledgement}
\vspace{-6pt}
This work is supported by the 
Flemish Government and Flanders Innovation \& Entrepreneurship (VLAIO) through the Flux50 projects InduFlexControl (HBC.2019.0113), and project
\href{https://www.improcap.eu/}{IMPROcap} (HBC.2022.0733).


\bibliographystyle{IEEEtran}
\bibliography{reference}

\end{document}